\documentclass[twocolumn,prl,preprintnumbers]{revtex4}
\usepackage{dcolumn}
\usepackage{bm}
\usepackage{ulem}
\usepackage{epsfig}
\usepackage{epstopdf}
\usepackage{amsmath}
\usepackage{graphicx}
\usepackage{mathrsfs}
\usepackage{amssymb}
\usepackage{amsfonts}
\usepackage{hyperref}
\usepackage{color,soul}
\usepackage{float}

\newcommand{\Tr}{{\rm Tr}}
\newcommand{\ket}[1]{|#1\rangle}               
\newcommand{\bra}[1]{\langle #1|}              

\begin{document}
\setlength{\textfloatsep}{2pt}
\setlength{\intextsep}{2pt}
\title{Uncertainty conservation relations: theory and experiment}

\author{Hengyan Wang$^{1^*}$}
\author{Zhihao Ma$^{2^*}$}
\author{Shengjun Wu$^{3^*}$}
\author{Wenqiang Zheng$^{4}$}
\author{Zhu Cao$^{5}$}
\author{Zhihua Chen$^{4}$}
\author{Zhaokai Li$^{1,6}$}
\author{Shao-Ming Fei$^{7,8}$}
\author{Xinhua Peng$^{1,6^\dag}$}
\author{Vlatko Vedral$^{9,10^\ddag}$}
\author{Jiangfeng Du$^{1,6^\S}$}

\affiliation{$^{1}$ CAS Key Laboratory of Microscale Magnetic Resonance and Department of Modern Physics, University of Science and Technology of China, Hefei 230026, China\\
$^{2}$ Department of Mathematics, Shanghai Jiaotong University, Shanghai 200240, China\\
$^{3}$ Kuang Yaming Honors School, Nanjing University, Nanjing, Jiangsu 210023, China\\
$^{4}$ Department of Applied Physics, Zhejiang University of Technology, Hangzhou, Zhejiang 310023, China\\
$^{5}$ Center for Quantum Information, Institute for Interdisciplinary Information Sciences, Tsinghua University, Beijing 100084 China\\
$^{6}$ Synergetic Innovation Center of Quantum Information and Quantum Physics, University of Science and Technology of China, Hefei 230026, China\\
$^{7}$ School of Mathematical Sciences, Capital Normal University, Beijing 100048, China\\
$^{8}$ Max-Planck-Institute for Mathematics in the Sciences, 04103 Leipzig, Germany\\
$^{9}$ Atomic and Laser Physics, University of Oxford, Clarendon Laboratory, Parks Road, Oxford, OX1 3PU, United Kingdom\\
$^{10}$ Centre for Quantum Technologies, National University of Singapore, 3 Science Drive 2, 117543 Singapore\\
$^*$These authors contributed equally to this work.\\
E-mail: $^\dag$ xhpeng@ustc.edu.cn, $^\ddag$ phyvv@nus.edu.sg, $^\S$ djf@ustc.edu.cn}

\maketitle

{\bf As a very fundamental principle in quantum physics, uncertainty principle has been studied intensively via various uncertainty inequalities.
Based on the information measure introduced by Brukner and Zeilinger in [Phys. Rev. Lett. 83, 3354 (1999)], we derive an uncertainty conservation relation
in the presence of quantum memory. We show that the sum of measurement uncertainties over a complete set of mutually unbiased bases
on a subsystem is equal to a total uncertainty determined by the initial bipartite state.
Hence for a given bipartite system the uncertainty is conserved and the uncertainty relation is given by an equality.
When the total uncertainty vanishes, all the uncertainties related to every mutually unbiased base measurement are zero, which is substantially different from
the uncertainty relation given by Berta et. al. [Nat. Phys. 6, 659 (2010)] where an uncertainty inequality is presented.
By directly measuring the corresponding quantum mechanical observables on a 5-qubit spin system, we experimentally verify the uncertainty conservation relation
in the presence of quantum memory without quantum state tomography.}

\noindent{\sf INTRODUCTION}

The uncertainty principle is one of the most important principle in quantum physics. It implies the impossibility of simultaneously determining the definite values of incompatible observables. The more precisely an observable is determined, the less precisely a complementary observable can be known. Based on the distributions of measurement outcomes, the quantum
uncertainty relations can be described in various ways, see for instance\cite{Heisenberg,Robertson,sch30,y6,y11,wym09,pnas110-6742,FS,PZ,Vallone,Maccone,rmp86-1261}.

The uncertainty principle was first formulated via the standard deviation of a pair of complementary observables, known as the Heisenberg's uncertainty principle \cite{Heisenberg} $\Delta x \Delta p \geq \hbar /2$ for the coordinate $x$ and the momentum $p$ in an infinite dimensional Hilbert space. Later the Robertson-Schr{\"o}dinger uncertainty inequality \cite{Robertson,sch30} presented an uncertainty relation for two arbitrary observables in a finite dimensional Hilbert space. Instead of the standard deviation of observables, the uncertainty principle can also be elegantly formulated in terms of entropies related to measurement bases.  When a quantum system is projected onto a certain basis $\{ \left| i_{\theta} \right\rangle\, | i=1,2,\cdots, d \}$, the uncertainty of the measurement results has been characterized by the Shannon entropy $H_{\theta}=\sum_{i=1}^d  -p_i \log _2 p_i$, where $p_i$ is the probability to obtain the $i$-th basis state $\left| i_{\theta} \right\rangle$. The larger the Shannon entropy $H_{\theta}$, the more uncertain the measurement results. In terms of the Shannon entropies of the measurement results, the uncertainty principle can be formulated as \cite{y11} :$H_{\theta} +H_{\tau} \geq \log_{2} \frac{1}{c}$. Here the term $c$ quantifies the degree of complementarity of the two measurements $\theta$ and $\tau$.

The above uncertainty relations only concern a single quantum system. By taking the entanglement with a memory system into account \cite{Horodecki09}, an entropic uncertainty relation in the presence of quantum memory has been investigated in Ref. \cite{np}. It has been shown that for a bipartite state $\rho_{AB}$, performing measurements on one of the subsystems $A$ gives rise to the following relation,
\begin{equation}\label{np}
S (\theta | B) + S (\tau | B ) \geq \log_{2} \frac{1}{c} + S(A|B),
\end{equation}
where $S(A|B)$ and $S(\theta | B)$ ($S (\tau | B )$) denote, respectively, the conditional von Neumann entropies of the initial bipartite state $\rho_{AB}$ and the final bipartite state $\rho_{\theta B}$ ($\rho_{\tau B}$) after the measurement in the basis $\{\left|i_{\theta} \right\rangle\}$  ($\{\left| i_{\tau} \right\rangle\}$).
This uncertainty relation was further extended to the smooth entropy case \cite{marco}, which was found to be applicable in quantum key distributions (QKD) \cite{marco2,Gehring}.

The entropy uncertainty relation in (\ref{np}) is an inequality, and concerns only measurements
on two arbitrary observables. In this work, we consider projective measurements based on mutually unbiased bases (MUBs) \cite{Bradler,mubs,Klappenecker,mubs2,mubs3,zhu}.
The MUB measurements are complementary to each other in the sense that any pair of bases are maximally unbiased. They are deeply connected to the Born's principle of complementarity \cite{Klappenecker} and closely related to the wave-particle duality \cite{Englert,brukner99}. A complete set of MUBs consists of at most $d+1$ observables, where $d$ is the dimension of the state space. For instance, there are three MUB observables in one qubit case, i.e., the Pauli operators $\sigma_x$, $\sigma_y$ and $\sigma_z$. Comparing with the incomplete case, the advantage of a complete set of MUB measurements is informatively complete \cite{mubs3,zhu} and meaningful in quantum information progressing \cite{Klappenecker}. It is therefore not surprising that a complete set of MUB measurements is crucial in entanglement detection \cite{Spengler12,Lu}. It was also proved that using a complete set of MUBs is much better than using two observables in QKD \cite{Bradler}, see references in \cite{mubs3,zhu} for other applications.

In Ref. \cite{wym09} an entropic uncertainty relation involving $d+1$ MUB measurements has been obtained in terms of von Neumann entropy. However, it only dealt with a single system (in this case the von Neumann entropy is just the Shannon entropy of the measurement probability distributions), and the uncertainty relation is given by an inequality.
In fact, the Shannon entropy is
a natural measure of our ignorance regarding the properties of a classical system, because in classical measurements the observation removes our ignorance about the state by revealing the properties of the system
which are considered to be pre-existing and independent of the observation. In contrast to classical measurements, one can not say that quantum measurements reveal a pre-existing property of a quantum system.
Therefore, the Shannon entropy could be thought of as ``conceptually" inadequate in quantum physics \cite{brukner01}. By taking into account that, for quantum systems the only features known before a measurement are the probabilities for various events to occur. In Ref. \cite{brukner99} the authors proposed a new measure of quantum information, which has significant physical meaning and various applications in quantum information processing such as quantum randomness, quantum state estimation, quantum teleportation and quantum metrology \cite{refm1,refm2,refm3,refm4,refm5,refm6,metrology}. Moreover, it has been shown that the sum of the individual measures of information for MUBs is invariant under the choice of the particular set of complementary observations and conserved if there is no information exchange with environments.

In this article, we adopt the information measure proposed  in Ref. \cite{brukner99} and consider the uncertainty relation in the presence of quantum memory.
Interestingly, we find that if we take a complete set of MUB measurements into account, we can obtain an uncertainty equality
that the sum of measurement uncertainties over all MUBs
on a subsystem in the presence of quantum memory is equal to a fixed quantity determined by the initial state, which gives rise to a
conservation relation of the uncertainties related to these MUB measurements.
We experimentally verify this uncertainty conservation relation by direct measuring uncertainties on a nuclear spin system.

\bigskip
\noindent{\sf RESULTS}

\smallskip
\noindent{\bf Theory}

Let ($p_1,p_2,...,p_d$) be the probabilities for the $d$ measurement outcomes. The lack of information about the $j$-th outcome
with respect to a single experimental trial is given by $p_j(1-p_j)$.
The total lack of information regarding all $d$ possible experimental outcomes is then given by
$\sum_{j=1}^{d} p_j(1-p_j)=1-\sum_{j=1}^{d} p_j^2$, which is minimal if one probability is equal to
a unity and maximal if all the probabilities are equal. In fact,
$1-\sum_{j=1}^{d} p_j^2$ is nothing but $1-\mathrm{Tr}(\rho^{2})$, where $\rho$ is the state after a quantum
(projective) measurement, the linear entropy of the measured state. Therefore,
the lack of information regarding all $d$ possible experimental outcomes
can be described by the linear entropy of a $d$-level quantum state $\rho$,
$S_{L}(\rho)= 1-\mathrm{Tr}(\rho^{2})$.
$S_{L}(\rho)$ ranges from $0$ (when $\rho$ is a pure state) to $(d-1)/d$ (when $\rho$ is maximally mixed).
Unlike that in Ref. \cite{brukner99}, here we do not introduce a normalization factor to have a range between $0$ and $\log_2 d$,
so the measure of uncertainty in terms of linear entropy does not have the unit of a ``bit". However, it quantifies uncertainty in a natural way:
an uncertainty of $0$ means the outcome is $100\%$ certain while an uncertainty approaching $1$ means the outcome is almost random.

For a bipartite state $\rho_{AB}$ in a $d \times D$ ($D\geq d$) dimensional composite Hilbert space, if system $A$ is projected on to the basis $\{\left|i_{\theta} \right\rangle | i=1,2,\cdots,d \}$,
the overall state of the composite system after the measurement on $A$ is given as
\begin{equation}
\rho_{\theta B} =
\sum_{i=1}^d \left| i_{\theta} \right\rangle _A \left\langle i_{\theta} \right|
\otimes _A\left\langle i_{\theta} \right|  \rho_{AB} \left| i_{\theta} \right\rangle _A  . \label{totalstateafterlocalmeasontheta}
\end{equation}
We can introduce the conditional linear entropy
\begin{equation}
\label{linear_entropy}
S_L ( \theta |B) := S_L(\rho_{\theta B}) -  S_L (\rho_{B}) =\mathrm{Tr} (\rho_{B}^2) -\mathrm{Tr} (\rho_{\theta B}^2)
\end{equation}
as a measure of the uncertainty about Alice's measurement result given Bob's state, where the reduced state $\rho_{B}=\Tr_A{(\rho_{\theta B})}=\Tr_A{(\rho_{A B})}$ is independent of the measurement basis.
It is straightforward to show that the conditional linear entropy $S_L ( \theta |B)$ is always non-negative.
As an example, suppose $\rho_{AB}$ is a maximally entangled pure state. Alice can perform a measurement on her system in any basis, the possible resulting states of Bob's system are orthogonal to each other, and each possible resulting state is in one-to-one correspondence to Alice's resulting state. Therefore, given Bob's state, Alice's measurement result can be determined with certainty. In this case $S_L ( \theta |B)$ vanishes.
If $\rho_{AB}=\sum_{i=1}^d \sqrt{\lambda_i} \left|i i \right\rangle$ is a partially entangled state written in its Schmidt bases,
after Alice measures her system in the Schmidt basis,
Bob's possible resulting states are orthogonal to each other and the Alice's measurement result is completely determined without uncertainty when Bob's state is given. This is also confirmed by the vanishing conditional entropy as $S_L(\rho_{\theta B}) = S_L (\rho_B) = \sum_i \lambda_i^2$. However, if Alice performs a measurement on a basis that is not the Schmidt basis,
the possible resulting states of Bob's system are not orthogonal and cannot be distinguished with certainty, and thus uncertainty of Alice's measurement result exists even when Bob's state is known. This fact is again confirmed by the observation that
the conditional linear entropy is strictly greater than zero in this case.
The conditional linear entropy is thus a good measure of the uncertainty about Alice's measurement result given Bob's state. It depends on the basis in which the measurement is performed in general.
When Alice tries to find a basis to perform the measurement on her system so that Bob will know her result with minimum uncertainty, then using another MUB to perform the measurement will result in Bob having a large uncertainty about Alice's result. However, the whole uncertainty running over all possible MUB measurements is conserved. This uncertainty relation is formulated in the following theorem (see proof
in Supplementary Materials).

{\bf Theorem 1.} For any density matrix $\rho_{AB}$ on a composite Hilbert space $H_{A}\otimes H_{B}$ of dimension $d\times D$,
we have the following uncertainty equality
\begin{equation}
\sum_{\theta=1}^{d+1} S_L ( \theta |B)  =  d \left(\mathrm{Tr}(\rho_{B}^{2})-\frac{1}{d} \mathrm{Tr}(\rho_{AB}^{2}) \right)   \label{UNCERTeq}
\end{equation}
when a complete set of $d+1$ MUBs exists for the $d$-dimensional Hilbert space $H_A$.

Theorem 1 shows that the total uncertainty related to the measurements over all $d+1$ MUBs of a subsystem is exactly given by a fixed quantity, $d \mathrm{Tr}(\rho_{B}^{2})- \mathrm{Tr}(\rho_{AB}^{2}) $, which is determined only by the initial bipartite state. This quantity is always nonnegative and can be viewed as the total measurement uncertainty of a subsystem, given the state of the other subsystem.
Different from inequality (\ref{np}) based on the von Neumann entropy, here we obtain the equality (\ref{UNCERTeq}). Note that this equality \eqref{UNCERTeq} is also
completely different from the one given in \cite{pra14} which is based on only ONE positive operator-valued measure consisting of uniformly all the measurement operators of $d+1$ MUBs (see \textit{Remarks} in Section I(C) of Supplementary Materials). In general, when there are only $M$ MUBs available or when we are only interested in certain $M$ MUBs, we always have the following uncertainty inequality,
\begin{equation}
\sum_{\theta=1}^{M} S_L ( \theta |B)  \geq  (M-1) \left(\mathrm{Tr}(\rho_{B}^{2})-\frac{1}{d} \mathrm{Tr}(\rho_{AB}^{2}) \right).
\label{UNCERTineq}
\end{equation}
With each additional MUB, the lower bound of total uncertainty is increased by a fixed amount $\mathrm{Tr}(\rho_{B}^{2})-\frac{1}{d} \mathrm{Tr}(\rho_{AB}^{2})$.

\begin{figure*}[h]
	\centering
	\includegraphics[width=14cm]{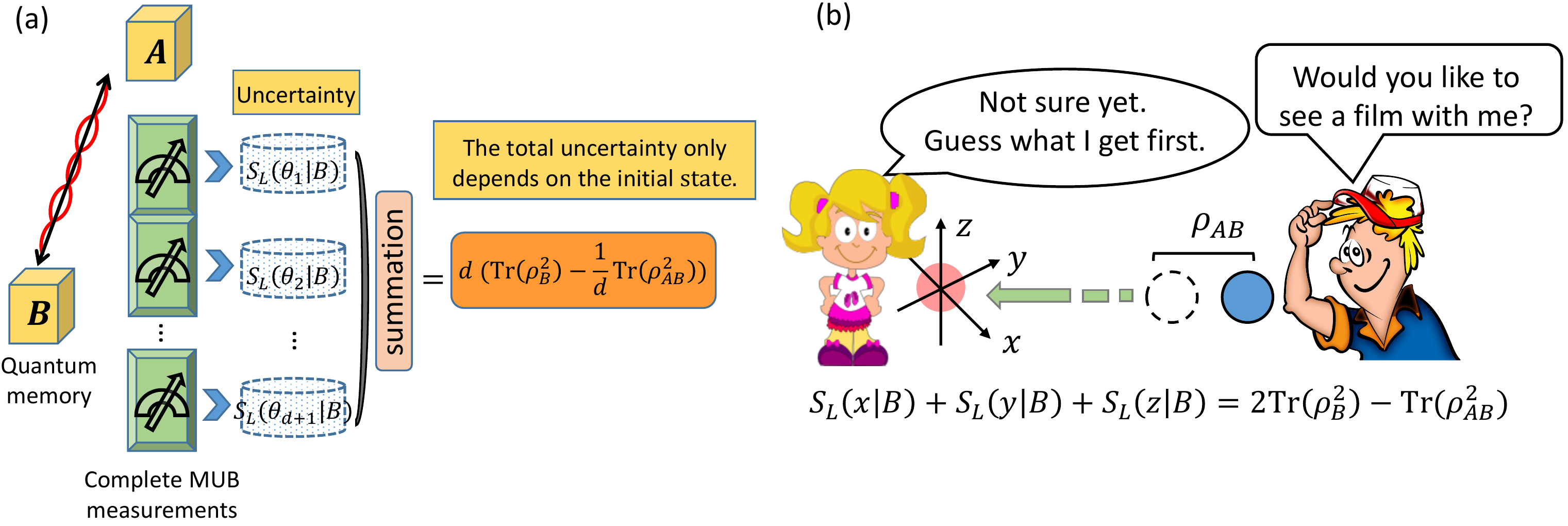}
	\caption{\label{fig1}
		\textbf{(a) Sketch of the proposal.
			(b) Illustration of implications of Theorem 1.} Alice chooses to measure one of the three Pauli matrices $\sigma_x$, $\sigma_y$ and $\sigma_z$ on qubit $ A $, and then informs Bob her choice and requests him to guess her measurement outcome.
		In order to guess Alice's measurement outcome with less uncertainty, Bob can entangle qubit $ A $ with a local qubit $ B $ before sending  qubit $ A $ to Alice, so that to
		minimize the uncertainty.}
\end{figure*}

To illustrate implications of Theorem 1, let us consider that Alice and Bob are both users of quantum technology. In order to make a hard decision on whether she should accept Bob's invitation to see a film,
Alice asks Bob to send her a qubit $A$. Alice can measure the qubit with the three (Pauli) observables $\sigma_x$, $\sigma_y$ and $\sigma_z$ at her choice.
After the measurement, Alice announces her choice of the observable, and Bob is supposed to guess the Alice's measurement results.
Alice would accept (deny) Bob's request if his guess is correct (wrong). Bob tries to gain Alice's acceptance by entangling the qubit $A$ with his local qubit $B$ in the preparation stage.
From the Theorem, we know that the sum of uncertainties (of Bob's guess at Alice's measurement results given the state of $B$) in three different cases is equal
to the quantity $Q=2\Tr (\rho_B^2) -\Tr (\rho_{AB}^2)$ that is completely determined by the initial state $\rho_{AB}$.
Bob can minimize the quantity $Q$ by preparing an EPR state, thus win Alice's acceptance with certainty, a result that can not be obtained from an uncertainty inequality like the one based on Shannon entropy, see Fig. \ref{fig1}.

On the practical side, Theorem 1 also provides possible applications in quantum random number generation, especially semi-self-testing quantum random number generators (QRNGs) which are more robust to device imperfections.
In a typical setup of a semi-self-testing QRNG\cite{Li2012Semi,cao15,cao16}, Alice and Bob share a quantum state $\rho_{AB}$, e.g., an EPR pair. If both parties are trusted, the measurement outcome of one party will be random to the other party when Alice measures in the computational basis and Bob measures in the diagonal basis. However, if one of the parties is corrupted, e.g., due to device imperfections, this scheme is broken. To show this, consider that one party switches to the same basis as the other party. A common solution is that each party randomly uses multiple basis, such as $\sigma_x$, $\sigma_y$, or $\sigma_z$ basis \cite{Li2012Semi,cao15,cao16}.
Now, we consider the following semi-self-testing scenario. Alice first chooses a reference frame, and randomly performs measurements in one of the MUB basis.
The reference frame is assumed to be reliably chosen, but Alice does not have a free will to randomly choose her measurement basis, i.e., the basis choice may be manipulated by an adversary who wishes to corrupt Alice's randomness, such as Bob.  Hence, the entropy of Alice's random outcomes with respect to Bob is the smallest entropy $S_L ( \theta |B)$ among all measurement choices.
To maximize this quantity, the Theorem shows that $S_L ( \theta |B)$  should be equal for all $\theta$s. Thus, the maximum entropy of a semi-self-testing QRNG is $[d \mathrm{Tr}(\rho_{B}^{2})- \mathrm{Tr}(\rho_{AB}^{2})]/(d+1)$.  This limit on semi-self-testing QRNGs also cannot be obtained from an uncertainty inequality. Finally, note that entanglement-based QRNGs considered here have a higher  randomness generation rate compared to prepare-and-measure QRNGs \cite{entanglebetter}, and cannot be analyzed using tools developed by Brukner and Zeilinger \cite{brukner99}. Further details will be published in a separate paper.

\smallskip
\noindent{\bf Experimental verification}

To experimentally investigate the uncertainty conservation, a two-qubit system $\rho_{AB}$ (i.e, a 4-dimensional quantum system), chosen as the test system, is prepared in the following states:
\begin{equation}
\begin{aligned}
\rho_{AB}(\alpha,x)&=x\ket{\psi_{\alpha}}\bra{\psi_{\alpha}}+\frac{1-x}{4}I_4,
\end{aligned}
\label{rhoAB}
\end{equation}
where $\ket{\psi_{\alpha}}=\cos(\alpha/2)\ket{01}-\sin(\alpha/2)\ket{10}$. The states are mixed states composed of one pure state with weight $x$ and the maximal mixed state with weight $(1-x)/4$.
The parameters $\alpha$ characterizes the entanglement of the pure part and $x$ characterizes the purity of the state.
When $\alpha=\pi/2$ and $x=1$, the bipartite state is a Bell states. The other three Bell states can be obtained by local unitary operations while the linear entropy remains invariant under such transformations.

\begin{figure*}[!ht]
	\centering
	\includegraphics[scale=0.45]{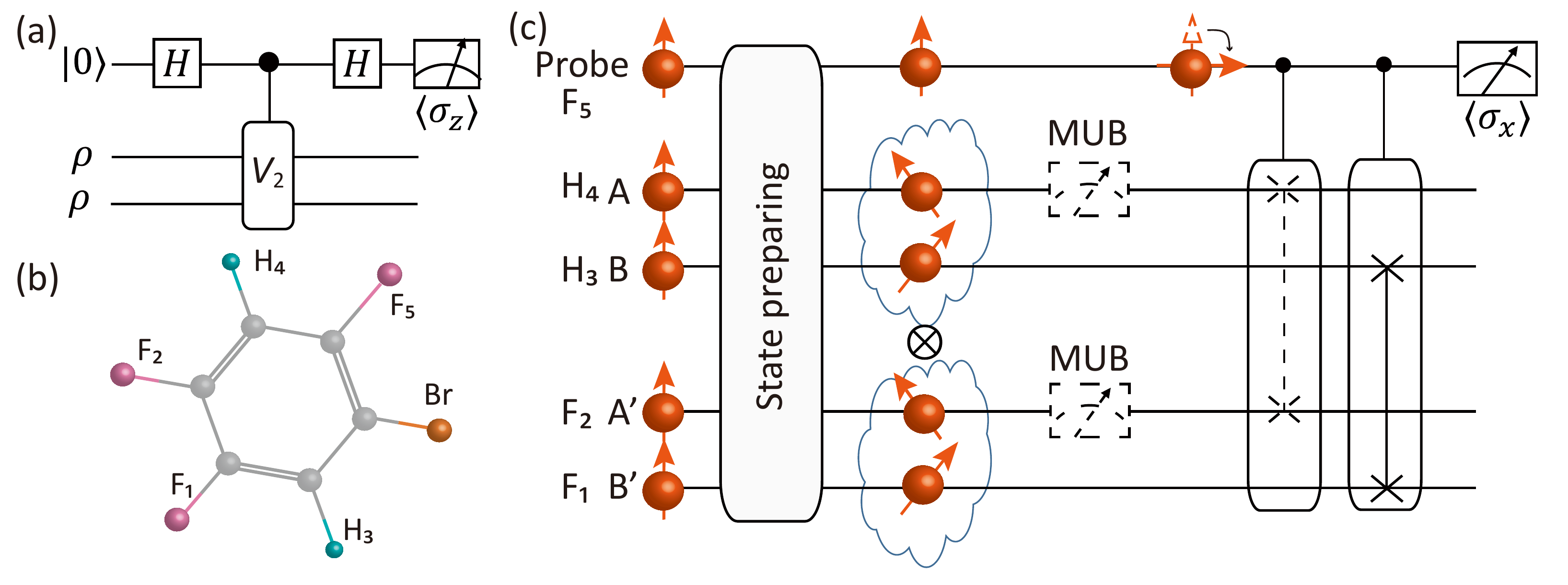}
	\caption{\label{schematic}
		\textbf{(a) Quantum circuit of directly measuring purity. \textbf{(b) Molecular structure for NMR quantum resgister.} (c) Experimental schematic for verifying the conversation relation.}
		MUB measurements are performed on subsystem $A$ to get the $\rho_{\theta B}$, and meanwhile the same process is
		applied on the mirror subsystem $A'$. Then controlled-SWAP gates are performed for the measurements
		of the linear entropies $S_L(\theta|B)$ in LHS of Eq. \eqref{UNCERTeq}. The purity information on the original
		state $\rho_{AB}$, i.e., the RHS of Eq. \eqref{UNCERTeq}, are obtained without MUB measurement in the dashed lines. The purities of the composite system $AB$: $\Tr(\rho_{A B}^2)$ and $\Tr(\rho_{\theta B}^2)$, are obtained by two controlled-
		SWAP gates $C_{swap}$ applying on subsystems $AA'$ and $BB'$ while only one $C_{swap}$ is operated on subsystem $BB'$ (denoted by the solid line) for the purity of subsystem $\Tr(\rho_B^2)$ and $\Tr(\rho_{B|\theta}^2)$, where
		$\rho_{B|\theta}=\Tr_A(\rho_{\theta B})$.}
\end{figure*}

The key part of the experiments is to measure the system's (conditional) linear entropy.
Similar to the measurement of von Neumann entropy in previous experiments \cite{npexpli,npexpPrevedel}, linear entropy can be indirectly measured by full quantum state tomography \cite{FST}. However it is inefficient for large-size quantum systems. Since the linear entropy is directly related to the purity $\mathrm{Tr}(\rho^2)$ that can be directly obtained by $\mathrm{Tr}(\rho^2)= \mathrm{Tr}\left( {V_2{\rho} \otimes {\rho}} \right)$ \cite{Trrho_2002} with a copy of $\rho$, as shown in Fig. \ref{schematic}(a), this allows us to employ an operational and direct way to experimentally verify the uncertainty conservation relation. Here the operator $V_2$ is just the SWAP operation, i.e., ${V_2}\left| {{\psi _1}{\psi _2}} \right\rangle = \left| {{\psi _2} {\psi _1}} \right\rangle$,  that exchanges the states of two subsystems. In the experiments, we introduced one ancillary probe qubit to perform the inteferometric measurement, directly obtaining all the required information of the purities from the probe qubit.

The experimental schematic is shown in Fig. \ref{schematic}(c), which consists of four steps: ($i$) to prepare initial state ${\rho _{in}} \propto  \sigma_z^{probe}  \otimes \rho _{AB} \otimes \rho_{A'B'}$, apart from the part of an identity matrix; ($ii$) to perform the MUB measurements under the basis $\{\left|i_{\alpha} \right\rangle | i=1,2,\cdots,d \}$; ($iii$) to carry out the controlled-SWAP circuit on the resulting state $\rho_{\theta B}$ after the MUB measurements; ($iv$) to measure the probe qubit to extract the purities $\mathrm{Tr}({\rho_{B}}^2)$ and $\mathrm{Tr}({\rho_{\theta B}}^2)$. Note that the measurements of linear entropies $\mathrm{Tr}({\rho_{AB}}^2)$ and $\mathrm{Tr}({\rho_{B}}^2)$ on the original state $\rho_{AB}$ are performed without step ($ii$).

Given its high control accuracy in multi-qubit system, the nuclear magnetic resonance (NMR) quantum processor is suitable for verifying the equation\cite{Ryan08}. We used the sample named 1-bromo-2,4,5-trifluorobenzene as a 5-qubit NMR quantum system which consists of two ${}^{1}$H spins and three ${}^{19}$F spins, dissolved in the liquid-crystal N-(4-methoxybenzylidene)-4-butylaniline (MBBA)(Fig. \ref{schematic}(b)).
Spins H$_3$ and H$_4$ are labeled as the bipartite system $\rho_{AB}$, spins F$_1$ and F$_2$ as the copy system ${\rho _{A'B'}}$, and spin F$_5$ as the probe qubit $\rho_{probe}$.
The effective Hamiltonian of the $5$-qubit system in rotating frame is
\begin{equation}
{H_\mathrm{NMR}} = \sum\limits_{i = 1}^5 {\pi {\nu _i}\sigma _z^i}  + \sum\limits_{1 \le j < k \le 5}^{} {\frac{\pi }{2}\left( {{J_{jk}} + 2{D_{jk}}} \right)\sigma _z^j\sigma _z^k},
\end{equation}
where $\sigma_z$ is the pauli operator, $\nu_i$ is the chemical shift of spin-$i$ and ${{J_{jk}} + 2{D_{jk}}}$ is the effective coupling constant of spin-$j$ and spin-$k$. The relevant parameters are shown in Supplementary Materials.

In order to prepare the initial state $\rho_{in}$, we first initialized the system to a labeled pseudo-pure state (LPPS) ${\rho _{LPPS}} =\frac{1}{32} {I_{32}}  + \varepsilon \sigma_z^{probe}  \otimes {\left| {0000} \right\rangle _{ABA'B'}}\langle 0000| $ from equilibrium state with selective-transition method \cite{pps2}, where $\varepsilon \approx 10^{-5}$ is the polarization and ${I_{32}}$ is the $32$-dimension identity matrix. In the following, we conventionally denote the state with the deviation density matrix \cite{12bitpps}, ignoring the identity matrix. Then $\rho_{in}$ was prepared by unitary and nonunitary operations from ${\rho _{LPPS}}$, where $ \rho_{A'B'} = \rho_{AB} =\rho_{AB}(\alpha,x)$.
We experimentally achieved the different weight $x$ by a rotation and a nonunitary gradient pulse (see Supplementary Materials).

The complete MUB measurements were implemented on subsystem $A$. For a 2-dimensional system, the simplest case of MUBs are
\begin{equation}\label{cmub}
\begin{array}{l}
M_0=\{\ket{0},\ket{1}\},~~~
M_1=\{\frac{\ket{0}+\ket{1}}{\sqrt{2}},\frac{\ket{0}-\ket{1}}{\sqrt{2}}\},\\[2mm]
M_2=\{\frac{\ket{0}+i\ket{1}}{\sqrt{2}},\frac{\ket{0}-i\ket{1}}{\sqrt{2}}\}.
\end{array}
\end{equation}
They are just the eigenvectors of Pauli operators $\sigma_x, \sigma_y$ and $\sigma_z$. The MUB measurements are emulated by gradient pulse fields in NMR \cite{Gzmeasure}.
All operations on $A'B'$ are exactly the same with system $AB$ and no correlation between them is generated.
Without interfering the unselected systems ($B$ and its copy $B'$), we realized the MUB measurements on system $A$ and its copy $A'$ by the gradient echo technology \cite{GradientEcho}.  
The readout process is similar with Fig. \ref{schematic}(a).
Firstly the probe qubit is transformed from $\sigma_z$ to $\sigma_x$ with one Hadamard gate for the interferometric measurements.  Then two controlled-SWAP gates (${C_{swap}}$) are applied to $AA'$ and $BB'$ respectively. Finally, through integrating the NMR spectra of the probe spin $\mathrm{F}_5$, we directly readout the value of its coherence term which just corresponds to the purity of bipartite system. The purity of subsystem $B$ is measured in a similar way while only one ${C_{swap}}$ gate is applied to $BB'$. Without the MUB measurements, we also measured the purity of the initial bipartite system $\rho_{AB}$ and subsystem $\rho_B$. By calculating the linear entropy, both sides of equation \eqref{UNCERTeq} are obtained without quantum state tomography.

\begin{figure*}[htp]
\centering
 \includegraphics[scale=0.5]{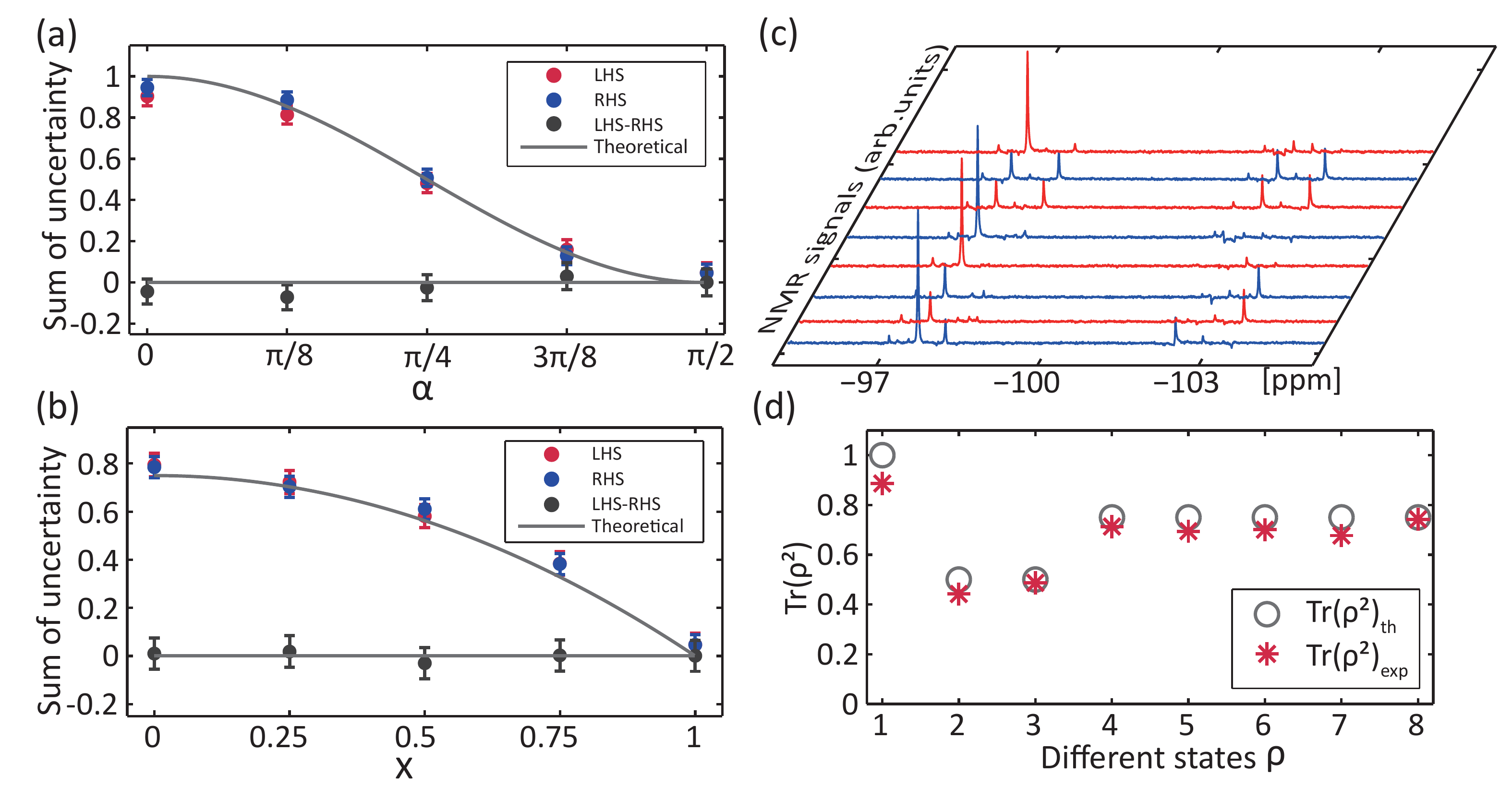}
\caption{{\bf Experimental results of verifying the uncertainty conservation with the input states \boldmath$\rho_{AB}(\alpha,1)$ (a)}  {\bf and} {\bf \boldmath$\rho_{AB}(\pi/2,x)$ (b).}
	Red and blue dots represent the measured values of LHS and RHS of equation (\ref{UNCERTeq}), respectively. The experimental data are rescaled by extracting the decoherence effect, shown by filled squares. The original data are shown in Supplementary Materials. The dark grey curves are theoretical expectations. The bars are plotted from the infidelity of readout process.
	\textbf{(c) Experimental NMR spectra of probe spin (F$_5$) for \boldmath$\rho_{AB}(\pi/4,1)$}. From bottom to up, the related resulting states after measurement are, respectively, $\rho_{AB}$, $\rho_{xB}$, $\rho_{yB}$, $\rho_{zB}$, $\rho_{B}$, $\rho_{B|x}$, $\rho_{B|y}$, $\rho_{B|z}$. \textbf{(d) Measured purities of the resulting states shown in (c).} Each purity is obtained from the sum of the integral values of the corresponding peak in (c).
}
\label{experimentresult}
\end{figure*}

The experimental results are shown in Fig. \ref{experimentresult}. As expected, the sum of uncertainties decreases to zero when the bipartite system is in maximally entangled state. With certain $\alpha$, lower purity corresponds to higher uncertainty.
From Fig. \ref{experimentresult}(a,b), we can see that the experimental results are in accord with theoretical expectations and the uncertainty conversation relation holds with high
precision. Fig. \ref{experimentresult}(c) shows the final NMR spectra of the probe qubit for one certain initial state $\rho_{AB}(\pi/4,1)$.
The sum of the integral values of all the peaks is read as the purity of the related state in our experiment, as shown in Fig. \ref{experimentresult}(d).

In our experiments, the unitary operations are realized by engineered quantum control pulses, which exploit the gradient ascent pulse engineering (GRAPE) algorithm \cite{GRAPE}. We numerically optimized all GRAPE pulses with considering $5\%$ ratio frequency (rf) field inhomogeneity, so that they are more robust in experiments. All GRAPE pulses used in the experiments have theoretical fidelities above $99.3\%$.
Numerical simulations show that the imperfection of GRAPE pulses causes infidelity of $2\%\sim4\%$ in the final states. Due to the short relaxation times of the liquid-crystal sample, the experiments suffer severe decoherence effect.
We numerically simulated the dynamical process and estimated the attenuation factors caused by decoherence effect in the experiments\cite{decoherence1,decoherence2}. Then we rescaled the experimental results. The details can be found in Methods and Supplementary Materials.
In the numerical simulations, we found that transverse relaxation time T$_2$ plays a leading role in the decoherence process, while
the longitude relaxation time T$_1$ has little influence.
The imperfection in preparing the labeled PPS also causes some errors. The highest unexpected peak in the LPPS NMR spectrum of spin F$_5$ is about $3\%$ intensity of the only expected peak.

\noindent{\sf CONCLUSION}

In conclusion, we have derived a novel entropic uncertainty conservation relation in bipartite systems with a quantum memory.
It has been shown that after a complete set of MUB measurements on one partite, the total uncertainty on the other partite is exactly given by the purities of the initial system and the memory. Substantially different from the previous uncertainty relations with inequalities, we in the first time presented an equality of uncertainty relation for the case with a quantum memory, which implies direct applications to quantum random number generation and quantum guessing games.
With the help of one mirror system of the measured system and one additional probe qubit, we have
provided the first experimental verification of this uncertainty conversation relation in an NMR quantum processor, where the experimental data of uncertainty quantities have been directly obtained by measuring the involved entropies without quantum state tomography. This method allows one to perform verification experiments in large quantum systems, and deal with the experimental data by standard statistical and information-theoretical methods.
These results may highlight further study on fundamental problems in quantum mechanics and the applications in quantum information processing such as quantum metrology and quantum computing.

\bigskip
\noindent{\sf MATERIALS AND METHODS}

\smallskip
\noindent\textbf{Proof of the theorem}

For a given bipartite quantum state $\rho_{AB}$ and a set of $M$ MUBs for system $A$, one can define an operator $\Gamma_{AB} \equiv I_A \otimes \rho_B + \frac{M-1}{d} \rho_{AB} - \sum_{\theta =1}^{M} \rho_{\theta B}$ on $\mathcal{H}_A \otimes \mathcal{H}_B$. In the supplement, we prove that the operator $\Gamma_{AB}$ vanishes when $M=d+1$ and is always nonnegative-definite when $M\leq d$. Therefore, $\Tr\{ \Gamma_{AB} \rho_{AB}\}$ vanishes when $M=d+1$ and is always nonnegative when $M\leq d$, these relations respectively yield the uncertainty equality (\ref{UNCERTeq}) and the uncertainty inequaltiy (\ref{UNCERTineq}).

\smallskip
\noindent\textbf{Direct measurement of \boldmath$\mathrm{Tr}(\rho^2)$}

In general case, we can measure $\mathrm{Tr}(\rho^k)$ with $k$ copies of $\rho$ by $\mathrm{Tr}(\rho^k)=\mathrm{Tr}(V_k\rho^{\otimes k})$, where $V_k$ is the shifter operator ${V_k}\left| {{\psi _1}{\psi _2}...{\psi _k}} \right\rangle  = \left| {{\psi _k} {\psi _1}...{\psi _{k - 1}}} \right\rangle $ \cite{Trrho_2002}. In the case of $k=2$, the shifter operator is just the SWAP operation, i.e., ${V_2}\left| {{\psi _1}{\psi _2}} \right\rangle = \left| {{\psi _2} {\psi _1}} \right\rangle$,  to exchange the states between two subsystems.
Then the purity $\Tr (\rho^2)$ can be directly obtained through measuring $\Tr(V_2\rho\otimes\rho)$.

Our initial state $\rho_{in} =\frac{1}{32} {I_{32}}  + \varepsilon \sigma_z^{probe}  \otimes \rho \otimes \rho$ is different from $| 0 \rangle \otimes \rho \otimes \rho$.  However, we will show the same quantum circuit measures the purity $\Tr (\rho_{AB}^2)$. The first Hadamard gate $H$ transforms the probe qubit state from $\sigma_z^{probe} $ to $\sigma_x^{probe} $, then the controlled-SWAP gate transforms the deviation density matrix of the whole system state into
\begin{widetext}
\begin{eqnarray}
&  &  {{C_{swap}}} (\sigma_x^{probe}  \otimes \rho \otimes \rho) {{C_{swap}}}^\dag \nonumber \\
& = &  (\left| 0 \right\rangle \left\langle 0 \right| \otimes I_2 + \left| 1 \right\rangle \left\langle 1 \right| \otimes V_2)[({ \left| 0 \right\rangle \left\langle 1 \right|  + \left| 1 \right\rangle \left\langle 0 \right| }) \otimes \rho \otimes \rho](\left| 0 \right\rangle \left\langle 0 \right| \otimes I_2 + \left| 1 \right\rangle \left\langle 1 \right| \otimes {V_2^\dag }) \nonumber \\
& = & \left| 0 \right\rangle \left\langle 1 \right| \otimes (\rho \otimes \rho) {V_2^\dag } + \left| 1 \right\rangle \left\langle 0 \right| \otimes V_2 (\rho \otimes \rho) \nonumber \\
& = &  \sigma_x^{probe} \otimes  V_2 (\rho \otimes \rho).
\end{eqnarray}
\end{widetext}
There is no dynamical and measurement effect on the part of identity density matrix. Therefore, the expectation value of $\sigma_x^{probe}$ gives the desired purity $\langle \sigma_x^{probe} \rangle = 2 \mathrm{Tr}(\rho^2) = 2 \mathrm{Tr}(V_2\rho\otimes\rho)$.



\smallskip
\noindent\textbf{Experimental MUB measurements}

In $2$-dimensional Hilbert space, the simplest three MUBs are
$M_0$, $M_1$ and $M_2$ given in (\ref{cmub}).
They just compose the eigenvectors of Pauli operators $\sigma_z, \sigma_x$ and $\sigma_y$. In NMR, such MUB projective measurements can be emulated using pulsed magnetic field gradients \cite{Gzmeasure}. A spatially dependent magnetic field gradient $G_z \, (T/m)$ along the $z$-axis causes a spatially dependent phase each coherence $\rho_{kl} (k \ne l)$ for a density matrix $\rho$. Thus after a gradient pulse with a suitable duration $t$, the density matrix averaged over the sample volume satisfies $\rho_{kl} =0$ for all  $k \ne l$ except for the zero quantum coherences for the homonuclear spins. Consequently, the coherences of $\rho$ by means of magnetic field gradients can be dephased, exactly as they would be by projective measurements of $\sigma_z$ on all the individual systems in the ensemble.
By gradient echo technique, i.e., a $\pi$ pulse to the other spins, this dephasing operation (i.e., projective measurement of $\sigma_z$) is selectively performed on some specific spins. It is in principle necessary to refocus all the evolutions under the internal
Hamiltonian during the gradient echo using $\pi$ pulses selective for single spins. However, this sequence can be simplified when we only care about the purity of the crashed state. For example, the MUBs (e.g., $\sigma_z$) on spin F$_2$ (A) and H$_4$ (A') are accomplished by
$P_{z}^{F_2,H_4} = G_z - [\pi]_{x}^{F_1,H_3,F_5} -G_z - [\pi]_{-x}^{F_1,H_3,F_5}$, where the evolutions under internal effective coupling Hamiltonian related to three qubits F$_1$, H$_3$ and F$_5$ are reserved during the pulse sequence. However, these undesired evolutions will lead to an error of less than $2[1-\cos(2 \Delta \theta)] \approx 0.065$ on the purity measurement, mainly determined by the different evolutions on qubits F$_1$, H$_3$ due to the different effective coupling constants $J_{F_1,F_5} + 2 D_{F_1,F_5}$ and $J_{H_3,F_5} + 2 D_{H_3,F_5}$. Here $\Delta \theta = \pi [(J_{F_1,F_5} + 2 D_{F_1,F_5} )- (J_{H_3,F_5} + 2 D_{H_3,F_5}) ] t_{G_z} /2$ with the duration $t_{G_z}$ of pulsed magnetic field gradient $G_z$.

To dephase the specific spins in the same way as would projective measurements of $\sigma_x $ and $\sigma_y$, one first selectively rotates the spins to the $z$ axis with $[ \pi /2]_{-y}$ or $[ \pi /2]_{x}$ rotations, then performs the projective measurement of $\sigma_z$, finally returns back to the base of $\sigma_x$ and $\sigma_y$, e.g., $P_{x}^{F_2,H_4} =  [ \pi /2]_{-y}^{F_2,H_4} - P_{z}^{F_2,H_4} - [ \pi /2]_{y}^{F_2,H_4} $ and $P_{y}^{F_2,H_4} =  [ \pi /2]_{x}^{F_2,H_4} - P_{z}^{F_2,H_4} - [ \pi /2]_{-x}^{F_2,H_4} $.

\smallskip
\noindent\textbf{The attenuation factors caused by decoherence effect}

To avoid the error accumulation and alleviate the influence of the decoherence, we used high-fidelity GRAPE pulses to implement the quantum circuit in the experiments.
The experiments for pure states contain $8$ GRAPE pules with the total duration of about $74ms$, while for mixed states, we used $9\sim 11$ GRAPE pulses with total durations of $67ms\sim85ms$.
However, experimental results are severely effected due to the decoherence due to the long running times comparing to the T$_2$ relaxation times of nuclear spins.
Therefore, we numerically simulated the experiments, extracted the attenuation factors caused by decoherence effect and then rescaled the experiment results for verifying the uncertainty conversation relation in Fig. \ref{experimentresult}(a) and Fig. \ref{experimentresult}(b). The original data without rescaling are also shown in Fig. 4(a) in Supplementary Materials.

In our experiments, the environmental noises are suitably modeled as Markovian, and  the evolution of the system is given by the Lindblad master equation
\begin{equation}
\label{Lindblad}
\dot \rho   =  - i[H_S + H_C(t),\rho ] +  \sum\limits_\alpha  {\left( {2{L_\alpha }\rho L_\alpha ^\dag  - L_\alpha ^\dag {L_\alpha }\rho  - \rho L_\alpha ^\dag {L_\alpha }}  \right)} ,
\end{equation}
where $H_S$ is the system Hamiltonian, $H_C(t)$ is the time-dependent external control Hamiltonian, and the operators ${{L_\alpha }}$ are Lindblad operators representing the coupling with the environment. The experiments are mainly effected by the dephasing process, and therefore ${{L_\alpha }} = \sqrt{\frac{\gamma_{\alpha}}{2}} \sigma_z^{\alpha}, (\alpha = 1,..., 5)$ and $\gamma_{\alpha} = 1/ T_{2 \alpha}$ and
\begin{equation}
\label{Lindblad2}
\dot \rho   =  - i[H_S + H_C(t),\rho ] +  \sum\limits_\alpha \gamma_{\alpha} {\left( \sigma_z^{\alpha} \rho \sigma_z^{\alpha}  - \rho  \right)} .
\end{equation}
In the experiments, the high-fidelity GRAPE pulses we used are kinds of shaped pulses consisting of thousands of slices with a constant $H_C(t_i)$ the duration of each slice, where $\delta t$ is $ 2\sim 25us$. Consequently, for each slice, the state of the system can be approximately given by
\begin{equation}
\begin{array}{l}
\rho(t_{i+1})  = \rho(t_i+\delta t)\\
=\displaystyle \sum_{\alpha=1}^5  \sum_{i=0}^1 E_i^{\alpha} e^{-i [H_S + H_C(t_{i+1})] \delta t } \rho(t_i) e^{i [H_S + H_C(t_{i+1})] \delta t } E_i^{\alpha \dagger},
\end{array}
\end{equation}
where the Kraus operations
\begin{eqnarray}
E_0^{\alpha} = \sqrt{\lambda_{\alpha}} \bf{1},   &  E_1^{\alpha} = \sqrt{1- \lambda_{\alpha}} \sigma_z^{\alpha}
\end{eqnarray}
with $\lambda_{\alpha} = (1 + e^{- \gamma_{\alpha} \delta t})/2$.

\bigskip
\noindent{\bf Acknowledgments}\\
We thank Marcin Paw{\l}owski, Simone Severini and Dawei Lu for helpfull discussions.
	This work was supported by
	National Key Basic Research Program of China (Grant No. 2013CB921800 and 2014CB848700),
	the National Science Fund for Distinguished Young Scholars (11425523),
	National Natural Science Foundation of China (Grants No. 11375167, 11661161018, 11227901, 11275131, 11571313, 11475084, 11605153, 11575173, 11675113), Strategic Priority Research Program (B) of the CAS (Grant No. XDB01030400), The National Key R\&D Plan of China (Grant No. 2016YFA0301801).

\noindent{\bf Author contributions}\\
Z. Ma., S. Wu. and X. Peng initiated the project. Z. Ma, S. Wu and Z. Chen proved the theorem. H. Wang, W. Zheng, Z. Chen and Z. Cao performed numerical analysis. Z. Cao, S. Fei and V. Vedral discussed and analyzed potential applications.
	X. Peng and J. Du supervised the experiment.
	X. Peng and H. Wang designed the experimental proposal.
	H. Wang and W. Zheng performed the experiment and analyzed the data.
	Z. Li and H. Wang tested the sample.
	All authors discussed the results, wrote the draft and reviewed the manuscript.

\noindent Correspondence and requests for materials should be addressed to X. P. (xhpeng@ustc.edu.cn), V. Vedral (phyvv@nus.edu.sg) or J. D. (djf@ustc.edu.cn).


\noindent{\bf Competing financial interests} The authors declare no competing financial interests.

\end{document}


\title{Supplementary Materials for \\
Uncertainty conservation relations: theory and experiment}

\author{Hengyan Wang}
\email{These authors contributed equally to this work.}
\affiliation{CAS Key Laboratory of Microscale Magnetic Resonance and Department of Modern Physics, University of Science and Technology of China, Hefei 230026, China}

\author{Zhihao Ma}
\email{These authors contributed equally to this work.}
\affiliation{Department of Mathematics, Shanghai Jiaotong University, Shanghai 200240, China}

\author{Shengjun Wu}
\email{These authors contributed equally to this work.}
\affiliation{Kuang Yaming Honors School, Nanjing University, Nanjing, Jiangsu 210023, China}

\author{Wenqiang Zheng}
\affiliation{Department of Applied Physics, Zhejiang University of Technology, Hangzhou, Zhejiang 310023, China}

\author{Zhu Cao}
\affiliation{Center for Quantum Information, Institute for Interdisciplinary Information Sciences, Tsinghua University, Beijing 100084 China}

\author{Zhihua Chen}
\affiliation{Department of Applied Physics, Zhejiang University of Technology, Hangzhou, Zhejiang 310023, China}

\author{Zhaokai Li}
\affiliation{CAS Key Laboratory of Microscale Magnetic Resonance and Department of Modern Physics, University of Science and Technology of China, Hefei 230026, China}
\affiliation{Synergetic Innovation Center of Quantum Information and Quantum Physics, University of Science and Technology of China, Hefei 230026, China}

\author{Shao-Ming Fei}
\affiliation{School of Mathematical Sciences, Capital Normal University, Beijing 100048, China}
\affiliation{Max-Planck-Institute for Mathematics in the Sciences, 04103 Leipzig, Germany}

\author{Xinhua Peng}
\email{xhpeng@ustc.edu.cn}
\affiliation{CAS Key Laboratory of Microscale Magnetic Resonance and Department of Modern Physics, University of Science and Technology of China, Hefei 230026, China}
\affiliation{Synergetic Innovation Center of Quantum Information and Quantum Physics, University of Science and Technology of China, Hefei 230026, China}

\author{Vlatko Vedral}
\email{phyvv@nus.edu.sg}
\affiliation{Atomic and Laser Physics, University of Oxford, Clarendon Laboratory, Parks Road, Oxford, OX1 3PU, United Kingdom}
\affiliation{Centre for Quantum Technologies, National University of Singapore, 3 Science Drive 2, 117543 Singapore}

\author{Jiangfeng Du}
\email{djf@ustc.edu.cn}
\affiliation{CAS Key Laboratory of Microscale Magnetic Resonance and Department of Modern Physics, University of Science and Technology of China, Hefei 230026, China}
\affiliation{Synergetic Innovation Center of Quantum Information and Quantum Physics, University of Science and Technology of China, Hefei 230026, China}

\maketitle

\section{Proof of the theorem}

\subsection{Mutually unbiased bases and bipartite basis states}

In a $d$-dimensional Hilbert space $\mathcal{H}$, let $\{\left| i_{\theta} \right\rangle | i=1,\cdots,d  \}$ denote a basis labeled by $\theta$. A set of $M$ such bases is called mutually unbiased if
\begin{equation}
|\left\langle i_{\theta} | j_{\tau} \right\rangle |^2 = \frac{1}{d}   \label{1}
\end{equation}
for any $i,j=1,2,\cdots,d$, and $\theta, \tau =1,\cdots, M$ with $\theta \neq \tau$.
When $d$ is a power of a prime number, a complete set of $d+1$ MUBs exists. When $d$ is an arbitrary integer number, the maximal number of MUBs is unknown. For example, when $d=6$, only $3$ MUBs have yet been found.  However, for any $d\geq 2$, there exist at least $3$ MUBs. For the purpose of this paper, we only assume that $M$ MUBs are available in a $d$-dimensional Hilbert space, where $M$ is less than or equal to the maximal number of MUBs that can exist.

In a composite Hilbert space $\mathcal{H} \otimes \mathcal{H}$ of two qudits, we construct the following
$M(d-1)+1$ states,
\begin{eqnarray}
|\Phi\rangle &=&\frac 1{\sqrt d}\sum_{i=1}^{d}|i_1\rangle_A\otimes |i_1\rangle_B^*,  \label{2} \\
|\phi_{\theta,k}\rangle &=& \frac 1{\sqrt d}\sum_{i=1}^{d}\omega^{k(i-1)}|i_\theta\rangle_A\otimes |i_\theta\rangle_B^* \label{3}
\end{eqnarray}
with $\omega=e^{2\pi i/d}$, $k=1,\ldots, d-1$ and $\theta=1,2,\ldots,M$. Here $|i_\theta\rangle^*$ denotes the complex conjugate of $|i_\theta\rangle$ with respect to the computational basis (which can be chosen as the first basis $\{|i_1\rangle\}$ without loss of generality).
It is straightforward to show the following proposition.

{\bf Proposition 1.}
The $M(d-1)+1$ bipartite states defined in Eqs. (\ref{2}, \ref{3}) are normalized and orthogonal to each other.

Therefore, these $M(d-1)+1$ states can be used for constructing a basis in the composite Hilbert space $\mathcal{H} \otimes \mathcal{H}$. Since there are at most $d^2$ orthogonal states in a $d^2$-dimensional space, one has $M(d-1)+1\leq d^2$, which implies $M\leq d+1$, i.e., there are at most $d+1$ MUBs for a $d$-dimensional Hilbert space. When $M=d+1$, i.e., a complete set of $d+1$ MUBs in a $d$-dimensional space is available, the states defined in Eqs. (\ref{2},\ref{3}) constitute a complete basis for the composite Hilbert space $\mathcal{H} \otimes \mathcal{H}$. On the other hand, when $M<d+1$, one can complete a basis of the composite Hilbert space by adding $p=(d-1)(d+1-M)$ additional orthonormal states $\{\left| \varphi_\alpha \right\rangle |\alpha =1,\cdots,p \}$. The projector onto the subspace spanned by these additional states is denoted by $\mathcal{P}$, i.e.,
\begin{equation}
\mathcal{P}  = I\otimes I - \left| \Phi \right\rangle \left\langle \Phi \right| -\sum_{\theta=1}^{M} \sum_{k=1}^{d-1} \left| \phi_{\theta,k} \right\rangle \left\langle \phi_{\theta,k} \right| =\sum_{\alpha=1}^{p} \left| \varphi_\alpha \right\rangle \left\langle \varphi_\alpha \right| .  \label{projectordef}
\end{equation}
It is obvious that $\mathcal{P} =0$ when $M=d+1$.

Let $T_2$ denote the partial transpose with respect to the computational basis of the second Hilbert space. One immediately has
\begin{equation}
(\left| \Phi \right\rangle \left\langle \Phi \right|)^{T_2} = \frac{1}{d} \sum_{i,j=1}^{d} \left| i_1 \right\rangle \left\langle j_1 \right| \otimes \left| j_1 \right\rangle \left\langle i_1 \right| .  \label{sumT1} \\
\end{equation}
As $\sum_{k=1}^{d-1} \omega^{k(i-j)}$ equals to $-1$ for $i\neq j$ and $d-1$ when $i=j$, it is not difficult to show
\begin{eqnarray}
 \left(\sum_{k=1}^{d-1} \left| \phi_{\theta,k} \right\rangle \left\langle \phi_{\theta,k} \right| \right)^{T_2} &=&  \sum_{i=1}^{d} \left| i_\theta \right\rangle \left\langle i_\theta \right| \otimes \left| i_\theta \right\rangle \left\langle i_\theta \right| \nonumber \\
 &-&\frac{1}{d}  \sum_{i,j=1}^{d} \left| i_\theta \right\rangle \left\langle j_\theta \right| \otimes \left| j_\theta \right\rangle \left\langle i_\theta \right|.
\label{sumT2}
\end{eqnarray}

\subsection{A useful nonnegative-definite operator}

Suppose $\rho_{AB}$ is a bipartite state on the composite Hilbert space $\mathcal{H}_A \otimes \mathcal{H}_B$ of dimension $d \times D$, and suppose $\{ \left| i_{\theta} \right\rangle_A \}$ is the $\theta$-th MUB in the $d$-dimensional Hilbert space $\mathcal{H}_A$, after system $A$ is projected onto the $\theta$-th MUB the overall
bipartite state is written as
\begin{equation}
\rho_{\theta B} =
\sum_{i=1}^d \left| i_{\theta} \right\rangle _A \left\langle i_{\theta} \right|
\otimes _A\left\langle i_{\theta} \right|  \rho_{AB} \left| i_{\theta} \right\rangle _A .  \label{totalstateafterlocalmeasonthetarepeat2}
\end{equation}
Given a bipartite state $\rho_{AB}$ and a set of $M$ MUBs in $\mathcal{H}_A$, we define an operator
\begin{equation}
\Gamma_{AB} \equiv I_A \otimes \rho_B + \frac{M-1}{d} \rho_{AB} - \sum_{\theta =1}^{M} \rho_{\theta B}   \label{Gammadef}
\end{equation}
on $\mathcal{H}_A \otimes \mathcal{H}_B$. This operator is Hermitian, and it has the following nice property.

{\bf Proposition 2.}
When $M=d+1$, the operator $\Gamma_{AB}$ vanishes: $\Gamma_{AB}=0$. When $M\leq d$, it is nonnegative-definite
\begin{equation}
\Gamma_{AB} \geq 0  .  \label{13}
\end{equation}

In order to prove proposition 2, we introduce an additional Hilbert space $\mathcal{H}_C$ of dimension $d$, and introduce a linear map $\mathcal{F}$ that maps operators on $\mathcal{H}_C$ to operators on $\mathcal{H}_A$, such that $\mathcal{F} (\left| i_1 \right\rangle _C \left\langle j_1 \right|) = \left| i_1 \right\rangle _A \left\langle j_1 \right|$ ($i,j=1,\cdots,d$). One can easily show that $\mathcal{F} (\left| i_\theta \right\rangle _C \left\langle j_\theta \right|) = \left| i_\theta \right\rangle _A \left\langle j_\theta \right|$ for $\theta =1,\cdots,M$. Let $\mathcal{F}^{-1}$ denote the inverse map, and let $\rho_{CB} \equiv \mathcal{F}^{-1}(\rho_{AB})$ denote the corresponding state on $\mathcal{H}_C \otimes \mathcal{H}_B$ with respect to the state $\rho_{AB}$ on $\mathcal{H}_A \otimes \mathcal{H}_B$.
Therefore, the map $\mathcal{F} : \rho_{CB} \rightarrow \mathcal{F}(\rho_{CB})$ can also be conveniently written via a partial trace over system $C$
\begin{eqnarray}
\mathcal{F}(\rho_{CB}) &=& \sum_{ij} Tr_C \left\{ \left[ \left| i_\theta \right\rangle _A \left\langle j_\theta \right| \otimes \left| j_\theta \right\rangle _C \left\langle i_\theta \right| \right] \rho_{CB} \right\} \nonumber \\
&=& \sum_{ij}  \left| i_\theta \right\rangle _A \left\langle j_\theta \right| \otimes  \left\langle i_\theta \right| \rho_{CB} \left| j_\theta \right\rangle  \label{transf} \\
&=& \rho_{AB}  \nonumber
\end{eqnarray}
for any $\theta \in \{1,\cdots,M\}$.
Similarly, $\rho_{\theta B}$ can be written as
\begin{equation}
\rho_{\theta B} = Tr_C \left\{ \left[\sum_{i=1}^d \left| i_{\theta} \right\rangle _A \left\langle i_{\theta} \right| \otimes \left| i_{\theta} \right\rangle _C \left\langle i_{\theta} \right| \right]
\rho_{CB}   \right\}   .
\end{equation}
Hence
\begin{eqnarray}
& &\sum_{\theta=1}^{M} (\frac{1}{d}\rho_{AB} -\rho_{\theta B}) \nonumber \\
&=& \sum_{\theta=1}^{M} Tr_C \left\{ \frac{1}{d}\sum_{ij} \left| i_{\theta} \right\rangle _A \left\langle j_{\theta} \right| \otimes \left| j_{\theta} \right\rangle _C \left\langle i_{\theta} \right| \rho_{CB} \right\}   \nonumber \\
& & - \sum_{\theta=1}^{M} Tr_C \left\{\sum_i \left| i_{\theta} \right\rangle _A \left\langle i_{\theta} \right| \otimes \left| i_{\theta} \right\rangle _C \left\langle i_{\theta} \right|  \rho_{CB} \right\}  \nonumber \\
&=& -Tr_C \left\{  \left(\sum_{\theta=1}^{M}  \sum_{k=1}^{d-1} \left| \phi_{\theta,k} \right\rangle _{AC} \left\langle \phi_{\theta,k} \right|  \right)^{T_C}  \rho_{CB} \right\} .
\label{dsum1}
\end{eqnarray}
We have used Eq. (\ref{sumT2}) to obtain the last equality. From Eqs. (\ref{sumT1}) and (\ref{transf}) with $\theta =1$, we also have
\begin{equation}
\frac{1}{d} \rho_{AB} =  Tr_C \left\{ (\left| \Phi \right\rangle _{AC} \left\langle \Phi \right|)^{T_C}  \rho_{CB} \right\}. \label{dsum2}
\end{equation}

From Eqs. (\ref{dsum1}) and (\ref{dsum2}) and the obvious relation $I_A \otimes \rho_B =Tr_C [ (I_a\otimes I_C) \rho_{CB}]$,
we can rewrite $\Gamma_{AB}$ as
\begin{eqnarray}
\Gamma_{AB} &=& I_A \otimes \rho_B - \frac{1}{d} \rho_{AB} +  \sum_{\theta =1}^{M} ( \frac{1}{d} \rho_{AB}- \rho_{\theta B} ) \nonumber \\
&=& Tr_C \left\{ \mathcal{P}_{AC}^{T_C}   \rho_{CB} \right\}.
\end{eqnarray}
Here the operator $\mathcal{P}_{AC}$ is the projector defined on $\mathcal{H}_A\otimes \mathcal{H}_C$ according to (\ref{projectordef}), i.e.,
\begin{eqnarray}
 \mathcal{P}_{AC} &=& I \otimes I - \left| \Phi \right\rangle \left\langle \Phi \right| -  \sum_{\theta=1}^{M}\sum_{k=1}^{d-1} \left| \phi_{\theta,k} \right\rangle  \left\langle \phi_{\theta,k} \right| \nonumber \\
 &=& \sum_{\alpha=1}^{p} \left| \varphi_\alpha \right\rangle _{AC} \left\langle \varphi_\alpha \right| .
\end{eqnarray}
When $M=d+1$, then $p=0$, the states $\left| \varphi_\alpha \right\rangle _{AC}$ in $\mathcal{H}_A\otimes \mathcal{H}_C$  do not exist, both $\mathcal{P}_{AC} $ and $\Gamma_{AB}$ vanish.
When $M\leq d$, we have
\begin{eqnarray}
\Gamma_{AB} &=& Tr_C \left\{ \sum_{\alpha=1}^{p} \left( \left| \varphi_\alpha \right\rangle _{AC} \left\langle \varphi_\alpha \right| \right)^{T_C}   \rho_{CB} \right\} \nonumber \\
&=& \sum_{\alpha=1}^{p}  Tr_C \left\{ \left( \left| \varphi_\alpha \right\rangle _{AC} \left\langle \varphi_\alpha \right| \right)^{T_C}   \rho_{CB} \right\} \nonumber \\
&=& \sum_{\alpha=1}^{p}  Tr_C \left\{ \left( \left| \varphi_\alpha \right\rangle _{AC}\right)^{T_C}    \rho_{CB} \left(  \left\langle \varphi_\alpha \right| _{AC} \right)^{T_C} \right\} . \nonumber
\end{eqnarray}
The last equality is due to the fact that operators on $\mathcal{H}_C$ can have cyclic permutations under the partial trace over $C$.
Let $\Theta_\alpha \equiv \left( \left| \varphi_\alpha \right\rangle _{AC}\right)^{T_C} \sqrt{\rho_{CB}}$, which are
operators on $\mathcal{H}_A \otimes \mathcal{H}_C \otimes \mathcal{H}_B$. Then we have
\begin{equation}
\Gamma_{AB}
= \sum_{\alpha=1}^{p}  Tr_C \left\{ \Theta_\alpha  (\Theta_\alpha)^\dagger \right\} .
\end{equation}
Since $\Theta_\alpha  (\Theta_\alpha)^\dagger$ are always nonnegative-definite operators on $\mathcal{H}_A \otimes \mathcal{H}_C \otimes \mathcal{H}_B$, the operators $Tr_C \left\{ \Theta_\alpha  (\Theta_\alpha)^\dagger \right\}$ are nonnegative-definite operators on $\mathcal{H}_A \otimes \mathcal{H}_B$, so is their sum. Therefore $\Gamma_{AB} \geq 0$. This completes the proof of proposition 2.

\subsection{Uncertainty equality and inequality}

According to proposition 2, $\Gamma_{AB}$ is a nonnegative-definite operator when $M\leq d$, and it vanishes when $M = d+1$.
Hence, for any nonnegative-definite operator $\Pi_{AB}$ on $\mathcal{H}_A \otimes \mathcal{H}_B$,
\begin{equation}
\Tr_{AB} (\Gamma_{AB} \Pi_{AB}) \geq 0
\label{15}
\end{equation}
when $M\leq d$, and the inequality becomes an equality when $M = d+1$.

Let $\Pi_{AB} = \rho_{AB}$, Eq. (\ref{15}) yields
\begin{eqnarray}
0 &\leq & \Tr_{AB} (\Gamma_{AB} \rho_{AB}) \nonumber \\
  & = & \Tr (\rho_B^2) +\frac{M-1}{d} \Tr(\rho_{AB}^2) - \sum_{\theta=1}^M \Tr(\rho_{\theta B}^2) .  \label{16}
\end{eqnarray}
Therefore,
\begin{equation}
- \sum_{\theta=1}^M \Tr(\rho_{\theta B}^2)  \geq - \Tr (\rho_B^2)  -\frac{M-1}{d} \Tr(\rho_{AB}^2) .
\end{equation}
Adding $M \Tr (\rho_B^2)$ to the above inequality, we immediately have
\begin{equation}
\sum_{\theta=1}^M (\Tr(\rho_B^2) -\Tr(\rho_{\theta B}^2) )  \geq (M-1) \left( \Tr (\rho_B^2)  -\frac{1}{d} \Tr(\rho_{AB}^2) \right) . \label{17}
\end{equation}
The inequality becomes an equality when $M=d+1$. Thus, the theorem in the main text has been proved.

{\it Remark}
The approach we admitted here is methodologically similar to the one used in
[Phys. Rev. Lett. 83, 3354 (1999)] (see also [Phys. Rev. A. 63, 022113(2001)]).
In \cite{pra14} an elegant uncertainty equality has also been presented based on ONE
positive operator-valued measure consisting uniformly all the measurement operators of $d+1$ MUBs.
As we take into account all the $d+1$ MUB projection measurements individually, our uncertainty relation is
completely different from the one given in \cite{pra14}. In fact, the uncertainty relation in \cite{pra14}
cannot be applied to our QRNG self-testing scenario, because the equality derived in \cite{pra14} holds only
when the system is measured in one of the MUBs with uniformly random probability.
While our equality holds no matter whether the MUB is uniformly chosen. Such a property is crucial in practical QRNG application,
as one needs to restrict the input randomness to choose a mutually unbiased basis.

\section{Quantum register}
\begin{figure}[here]
\centering
\includegraphics[scale=0.42]{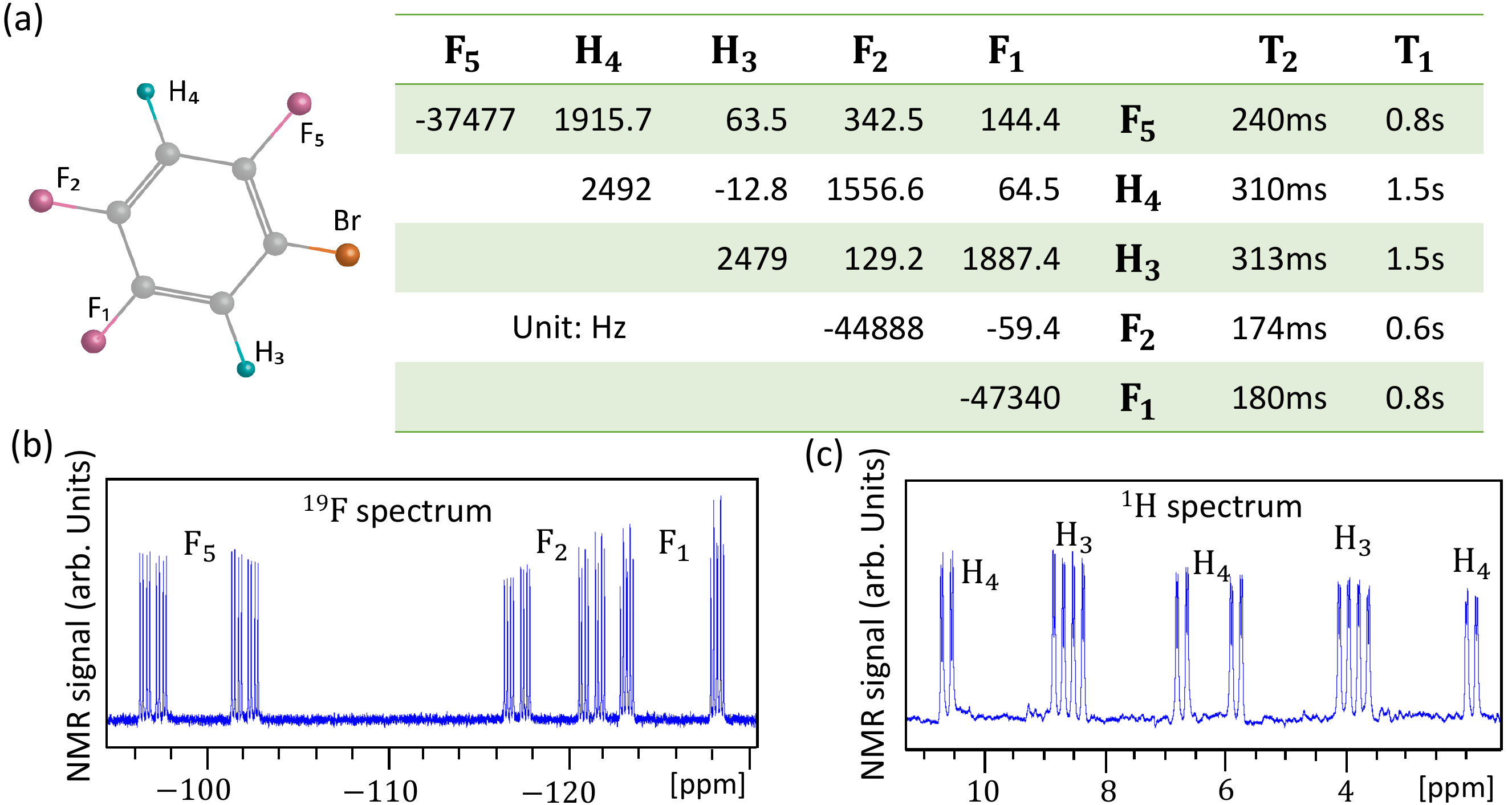}
\caption{
\textbf{(a)} Characteristics of the 1-bromo-2,4,5-trifluorobenzene molecule.  The chemical shifts and effective coupling
constants (in Hz) are on and above the diagonal in the table, respectively. The last two columns show the transversal relaxation times
and the longitudinal relaxation times of each nucleus.
\textbf{(b)} The thermal equilibrium ${}^{19}$F and ${}^{1}$H spectra.
}
\label{fig_SM_1}
\end{figure}

We used the sample named 1-bromo-2,4,5-trifluorobenzene as a 5-qubit NMR quantum system which consists of two ${}^{1}$H spins and three ${}^{19}$F spins, dissolved in the liquid crystal N-(4-methoxybenzylidene)-4-butylaniline. The structure of the molecule is shown in Fig. \ref{fig_SM_1}.
Due to the partial average effect in the liquid crystal solution,  the direct dipole-dipole interaction will be scaled down by the order parameter \cite{Dong97}. In our sample, the partial average effect makes the direct dipole-dipole couplings of homonuclear spins much smaller than the difference between the chemical shifts of related nuclear spins, all the dipole-dipole Hamiltonians are reduced to the form of ${\sigma_z^i}{\sigma_z^j}$. Therefore,
the effective Hamiltonian of the $5$-qubit system in rotating frame is
\begin{equation}
{H_\mathrm{NMR}} = \sum\limits_{i = 1}^5 {\pi {\nu _i}\sigma _z^i}  + \sum\limits_{1 \le j < k \le 5}^{} {\frac{\pi }{2}\left( {{J_{jk}} + 2{D_{jk}}} \right)\sigma _z^j\sigma _z^k},
\end{equation}
where $\sigma_z$ is the pauli operator, $\nu_i$ is the chemical shift of spin-$i$ and ${{J_{jk}} + 2{D_{jk}}}$ is the effective coupling constant of spin-$j$ and spin-$k$. The relevant parameters are shown in Fig. \ref{fig_SM_1}.

\section{Initial state preparation}
We first initialized the quantum register into a labeled pseudo-pure state (LPPS) $\rho _{LPPS}= \varepsilon\sigma_z^{probe}\otimes {\left| {0000} \right\rangle _{ABA'B'}}\langle 0000| + 1/32 {I_{32}}$ from the equilibrium state where $\varepsilon$ is the polarization about $10^{-5}$. We just neglected the identity part because unitary and non-unitary operations all have no influence on it.
We applied $30$ unitary operators with the form $U=e^{-i\beta_{ij} \mathcal{V}_{ij}}$ to redistribute the populations between energy levels $i$ and $j$. Here $\mathcal{V}_{ij}$ is the single quantum transition operator
between levels $i$ and $j$, that is, a $32 \times 32$ matrix whose elements are zero except for two elements $\mathcal{V}_{ij}(i,j)=\mathcal{V}_{ij}(j,i)=1/2$. According to the desired distribution, the angle $\beta_{ij}$ were numerically calculated. In this produce, the unitary operators generated undesired coherence terms that were eliminated by a following field gradient pulse $Gz$. Accordingly, the LPPS $\rho _{LPPS}$ was prepared.

In the following we only consider the deviation part $\sigma_z^{probe}\otimes{\left| {0000} \right\rangle _{ABA'B'}}\langle 0000| $. The composite system was prepared into the state
\begin{equation}
\begin{aligned}
\sigma_z^{probe}\otimes\rho_{AB}(\alpha,x)\otimes\rho_{A'B'}(\alpha,x) = \sigma_z^{probe}\otimes(x\ket{\psi_{\alpha}}\bra{\psi_{\alpha}}+\frac{1-x}{4}I_4)\otimes(x\ket{\psi_{\alpha}}\bra{\psi_{\alpha}}+\frac{1-x}{4}I_4),
\end{aligned}
\end{equation}
from $\sigma_z^{probe}\otimes\ket{0000}\bra{0000}$, where $| \psi_{\alpha} \rangle =\cos(\alpha/2)\ket{01}-\sin{\alpha/2}\ket{10}$.
\begin{figure}[here]
\centering
\includegraphics[scale=0.32]{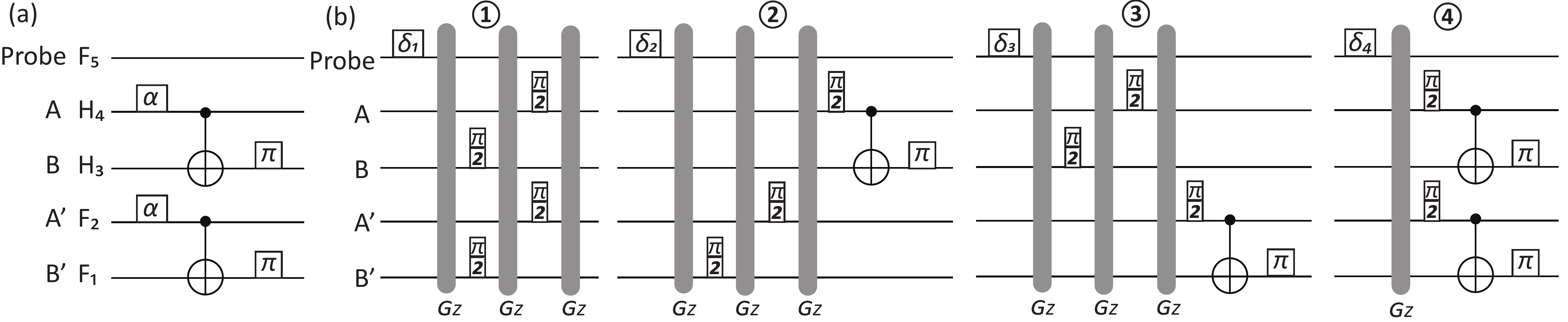}
\caption{
\textbf{(a)} Quantum circuit of preparing $\sigma_z^{probe}\otimes\rho_{AB}(\alpha,1)\otimes\rho_{A'B'}(\alpha,1)$ from LPPS. The rectangles represent single-qubit rotations with angles $\alpha$, $\pi/2$ and $\pi$ along $y$-axis. $Gz$ is the field gradient pulse along $z$-axis.
\textbf{(b)} Pulse sequences for preparing $ \sigma_z^{probe}\otimes\rho_{AB}(\pi/2,x)\otimes\rho_{A'B'}(\pi/2,x)$ with the temporal averaging method.
The rotation angles $\delta_{1,2,3,4}$ are $\arccos[(1-x)^2]$, $\arccos[x(1-x)]$, $\arccos[x(1-x)]$ and $\arccos(x^2)$, which are used to adjust the coefficients of $4$ parts in eg. (\ref{mix4parts}).
}
\label{fig_SM_2}
\end{figure}
In the case of $x=1$, the bipartite systems are entangled where the rotation angle $\alpha$ varies from $0$ to $\pi/2$ for producing different entanglement. $\alpha=0$ corresponds to separable state and $\alpha=\pi/2$ corresponds to the maximally entangled state. Setting $\alpha=\pi/2$ and scanning $x$ from $0$ to $1$ with interval $0.25$, we prepared the states with different mixedness.
In preparing  $ \sigma_z^{probe}\otimes\rho_{AB}(\pi/2,x)\otimes\rho_{A'B'}(\pi/2,x)$, we divided it into $4$ parts, i.e.,
\begin{equation}\label{mix4parts}
\begin{aligned}
&\sigma_z^{probe}\otimes(\frac{1-x}{4}I_2)\otimes(\frac{1-x}{4}I_2), \\ &\sigma_z^{probe}\otimes(\frac{1-x}{4}I_2)\otimes(x\ket{\psi_{\pi/2}}\bra{\psi_{\pi/2}}), \\ &\sigma_z^{probe}\otimes(x\ket{\psi_{\pi/2}}\bra{\psi_{\pi/2}})\otimes(\frac{1-x}{4}I_2), \\ &\sigma_z^{probe}\otimes(x\ket{\psi_{\pi/2}}\bra{\psi_{\pi/2}})\otimes(x\ket{\psi_{\pi/2}}\bra{\psi_{\pi/2}}), \end{aligned}
\end{equation}
and adopted the temporal averaging method \cite{TemporalAverage} to realize them. By summing the four output states, we obtained the desired state $ \sigma_z^{probe}\otimes\rho_{AB}(\pi/2,x)\otimes\rho_{A'B'}(\pi/2,x)$.
The pulse sequences for the state preparation are shown in Fig. \ref{fig_SM_2} b.

\section{Experimental procedure}

\begin{figure}[here]
\centering
\includegraphics[scale=0.4]{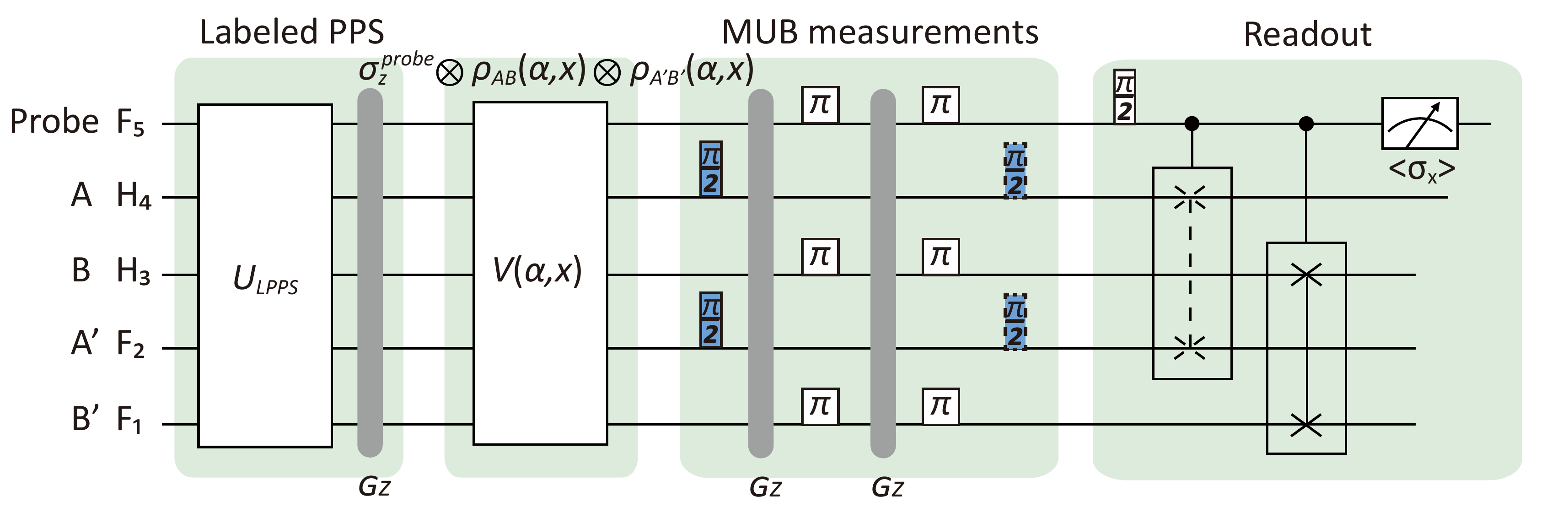}
\caption{
Experimental pulse sequence. The LPPS preparation contains one GRAPE pulse and one $z$-axis gradient pulse. The GRAPE pulse is used to redistribute the populations and the field gradient pulse is used to eliminate the undesired coherences.
The operation $V(\alpha,x)$ is to prepare the initial state $\sigma_z^{probe}\otimes\rho_{AB}(\alpha,x)\otimes\rho_{A'B'}(\alpha,x)$. Its specific form is shown in Fig. \ref{fig_SM_2}.
The solid blue rectangle is $\pi/2$ rotation along $(-y)$-axis in $x$-measurement and along $x$-axis in $y$-measurement. The dashed blue rectangles denotes $\pi/2$ rotations with the opposite phases to  those in solid blue rectangles. No need to operate them in $z$-measurement.
The readout part contains two ${C_{swap}}$ gates applying on subsystems $AA'$ and $BB'$. The dashed ${C_{swap}}(AA')$ is not required in reading out the purity of subsystem $B$.}
\label{fig_SM_3}
\end{figure}

The total experimental procedure is shown in Fig. \ref{fig_SM_3}, including the LPPS preparation, the initial state preparation, MUB measurements and the readout of the purities. When the measurement is along $y$-axis, the first and last rotations in the MUB measurement part should be substituted by rotations along $x$ and $-x$, and the angles remain also $\pi/2$. In the $z$-axis measurement case, these two rotations are omitted.

\section{Experimental results of directly measured purities}
\begin{figure}[here]
\centering
\includegraphics[scale=0.5]{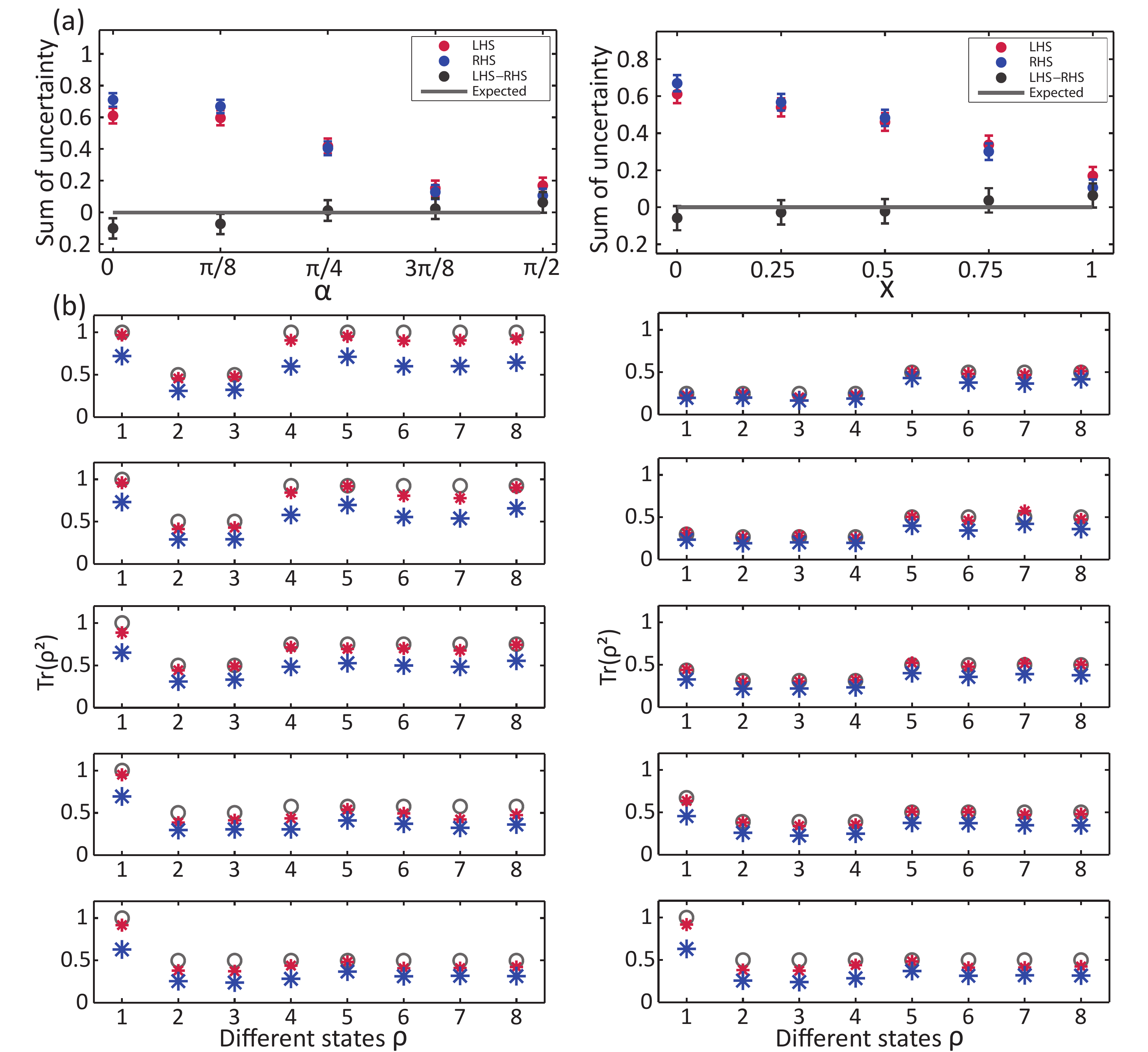}
\caption{
\textbf{(a)} The original results without rescaling of verifying the uncertainty conservation.
The bars are plotted from the infidelity of readout process.
\textbf{(b)} The directly measured purities.
he hollow dots are the expected purities. The blue asterisks are the directly measured purities and the red ones are the rescaled results.
The left five figures correspond to purities when the initial bipartite state is $\rho_{AB}(\alpha,1)$ where $\alpha=0,\pi/8,\pi/4,3\pi/8,\pi/2$ from top to bottom. And the right five figures correspond to $\rho_{AB}(\pi/2,x)$ where $x=0,0.25,0.5,0.75,1$ from top to bottom.
 In each figure, from left to right, they are $\Tr(\rho_{AB}^2)$, $\Tr(\rho_{xB}^2)$, $\Tr(\rho_{yB}^2)$, $\Tr(\rho_{zB}^2)$, $\Tr(\rho_{B}^2)$, $\Tr[(\rho_{B|x})^2]$, $\Tr[(\rho_{B|y})^2]$, $\Tr[(\rho_{B|z})^2]$, where $\rho_{\theta B}$ $(\theta = x,y,z)$ are the resulting state after the complete MUB measurements on subsystem $A$, and  $\rho_{B|\theta} = \Tr_A(\rho_{\theta B})$.
}
\label{fig_SM_4}
\end{figure}

The final experimental signals are influenced by decoherence effect. We simulated the decoherence effect and rescaled the final results. Fig. \ref{fig_SM_4}(a) shows the un-rescaled original results. From this we can see that even though the right and left sides of the uncertainty equation are all reduced, they almost equal to each other. Fig. \ref{fig_SM_4}(b) shows the directly measured the purities of different states required in the uncertainty conversation relation.

